\documentclass[]{aa}  

\usepackage{graphicx}
\usepackage[varg]{txfonts}
\usepackage{natbib}
\bibpunct{(}{)}{;}{a}{}{,} 
\begin{document}
   \title{Dark influences III. Structural characterization of minor mergers of dwarf galaxies with dark satellites}
   \titlerunning{Structural properties of dwarf galaxy minor merger remnants}
   \subtitle{}

  \author{T. K. Starkenburg\inst{1} 
         \and A. Helmi\inst{1} 
         \and L. V. Sales\inst{2}
         }

  \institute{Kapteyn Astronomical Institute, University of Groningen,
             P.O. Box 800, 9700 AV Groningen, The Netherlands\\
             \email{tjitske@astro.rug.nl}
             \and Department of Physics and Astronomy, University of California, Riverside, CA 92521, USA  
             }

  \date{Received date / Accepted date }

  \abstract{In the current concordance cosmology small halos are
    expected to be completely dark and can significantly perturb
    low-mass galaxies during minor merger interactions. These
    interactions may well contribute to the diversity of the dwarf
    galaxy population. Dwarf galaxies in the field are often observed
    to have peculiarities in their structure, morphology, and
    kinematics, as well as strong bursts of star formation without
    apparent cause.}  {We aim to characterise the signatures of minor
    mergers of dwarf galaxies with dark satellites to aid their
    observational identification.}  {We explored and quantified a variety
    of structural, morphological, and kinematic indicators of merging
    dwarf galaxies and their remnants using a suite of hydrodynamical
    simulations.}  {The most sensitive indicators of mergers with dark satellites are large
    asymmetries in the gaseous and stellar distributions, enhanced
    central surface brightness and starbursts, and velocity
    offsets and misalignments between the gas and stellar
    components. In general, merging systems span a wide range of values
    of the most commonly used indicators, while isolated objects tend to
    have more confined values. Interestingly, we find in our simulations that a
    significantly off-centred burst of star formation can pinpoint the
    location of the dark satellite. Observational systems with such
    characteristics are perhaps the most promising for unveiling the
    presence of the hitherto missing satellites.}{}

\maketitle
\section{Introduction}

In a Lambda cold dark matter ($\Lambda$CDM) universe the halo mass
function is scale-free: independently of their mass, halos have their
own system of substructures
\citep{vdBoschetal2005,vdBoschJiang2014}. Below a halo mass of $\sim
10^{9.5}\ M_{\sun}$ however star formation is expected to be largely
inhibited due to reionization, photo-ionization of the gas, and
possibly feedback \citep{Gnedin2000, Hoeftetal2006,
  Kaufmannetal2007, Okamotoetal2008, Gnedinetal2009, Lietal2010,
  Sawalaetal2013}. The galaxy mass function is thus not scale-free,
while the stellar mass-halo mass function is predicted to steepen
toward lower halo masses \citep{Behroozietal2013, Mosteretal2013, Garrison-Kimmeletal2014,
  Sawalaetal2015,
  KormendyFreeman2016}. Therefore, dwarf galaxy halos have significantly
lower baryon fractions and their satellites are expected to be
predominantly completely star-free, or dark \citep{Helmietal2012}.

Although the Hubble sequence \citep{Hubble1926} generally describes well the properties
of large galaxies, on the scale of dwarfs no clear classification
scheme exists. The simplest separation is given by the fact that star-forming dwarfs often show irregular morphologies, while those
quiescent have generally a spheroidal appearence. It is still not well
understood how these classes of objects are related \citep[see e.g.][]{Mateo1998ARA&A, TolstoyHillTosi2009ARA&A}.  Furthermore,
blue compact dwarfs (BCDs), and more generally starbursting dwarf
galaxies, have central regions that are very blue reflecting a
centrally concentrated young stellar population so bright that an
underlying, older population is not readily apparent
\citep[e.g.,][]{dePazetal2003,Paudeletal2015}. Just like dwarf
irregulars, BCDs can furthermore depict irregular morphologies and
kinematics, with star-formation regions far from the centre
\citep{Tayloretal1995,EktaChengalur2010,Lopez-Sanchez2010,Holwerdaetal2013VII,Lellietal2014c,KnapenCisternas2015}.
Off-centre bursts of star formation have also been observed in a
number of extremely metal-poor galaxies as well as large differences
in the average line-of-sight velocities between the HI gas and the stellar
component \citep{Filhoetal2013,Filhoetal2015}. As galaxy mass reduces,
it appears that a higher fraction of the systems are peculiar.

\begin{table*}[b!, 0.9\textwidth]
   \caption{\label{hostparm} Structural and numerical parameters for the host dwarf galaxies.}
         $$ 
         \begin{array}{llllllllllr}
            \hline
            \noalign{\smallskip}
              \mathrm{Model} & M_{\rm vir} & r_{\mathrm{vir}} & c & M_{\star} & R_d & \displaystyle\frac{z_0}{R_d} & f_g & \displaystyle\frac{R_g}{R_d} & \displaystyle \frac{M_{\rm sat}}{M_{\rm vir_{\rm main}}} \\
               & 10^{10} M_{\sun} & \mathrm{kpc} &  & 10^8 M_{\sun} & \mathrm{kpc} &  &  &  &  \\          
            \noalign{\smallskip}
            \hline
            \hline
            \noalign{\smallskip}
            \noalign{\smallskip}
            {\rm A} & 5.6 & 77 & 9 & 1.4 & 0.93 & 0.1 & 0.5 & 1;2;4 & 0.05;0.1;0.2 \\
            {\rm B} & 2.2 & 56 & 15 & 0.27 & 0.78 & 0.2 & 0.75 & 1 & 0.1;0.2 \\
            {\rm C} & 1.4 & 48 & 15;5 & 0.11 & 0.78;0.39 & 0.3 & 0.9 & 1;2 & 0.2 \\
            {\rm D} & 0.97 & 42 & 5;15 & 0.044 & 0.95;0.48 & 0.3;0.5 & 0.9 & 1;2 & 0.2 \\
            \noalign{\smallskip}
            \hline
            \hline
         \end{array}
     $$ 
   \end{table*}

We have recently postulated that this may be partly explained by dwarf
galaxies experiencing minor mergers with dark companions \citep{Helmietal2012}. In
\citet{SH15} and \citet{SHS16}, we show that such minor mergers
can significantly alter the morphological properties of dwarf
galaxies. The disturbances induced by dark objects are much more
dramatic on this scale because of the lower galaxy formation
efficiency (i.e. lower baryon fractions) in dwarfs compared to giant galaxies. One of the most
direct imprints in gas-rich dwarfs is a vast increase in star
formation: both in short bursts (during close passages of the
satellite) as well as sustained high star formation rates lasting several Gyrs. In
\citet{SHS16} we show that the general properties of our
simulated dwarf systems compare very well to a large sample of dwarf
irregular galaxies and blue compact dwarfs from the literature. 

In this paper we provide a quantitative characterization of the
morphological and kinematic properties of the dwarf systems during the
minor merger events and thereby facilitate a more detailed comparison
to observations. For the analysis we used morphological descriptions
that have been applied to characterise where galaxies lie along the Hubble sequence, to disentangle interacting from isolated
systems, and to describe the stellar distributions in major
mergers of $\sim L_{\star}$ or larger spiral galaxies, such as the
\emph{CAS} (concentration, asymmetry and smoothness) and \emph{GM}
(Gini coefficient and M20) indicators \citep[see e.g.][and references
therein]{Conseliceetal2000, Conselice2003, Abrahametal2003,
  Lotzetal2004}. These have also been applied to describe the stellar components of isolated irregular
dwarf galaxy samples \citep{Conselice2003,Lotzetal2004}, to
characterise the gas distribution in starbursting dwarf galaxies
\citep{Lellietal2014c} and in simulations of major mergers
\citep{Holwerdaetal2011III}, as well as a variety of observational
samples \citep{Holwerdaetal2011IV,Holwerdaetal2011I,Holwerdaetal2011II,Holwerdaetal2011V,Holwerdaetal2012VI,Holwerdaetal2013VII,
  Holwerdaetal2014}.

This paper is organised as follows. The hydrodynamical simulations are
described concisely in Sect. \ref{Method}, while in Sect. \ref{Tidal}
we focus on one specific simulated system and highlight key tidal
features as the merger takes place. In Sect. \ref{Quantitative} we
introduce the morphological and kinematic indicators used and compare
the results to some observational samples. We present a brief
comparison to two dwarf galaxies with peculiar properties, namely IC10 and NGC6822 in
Sect. \ref{Discussion}. The summary and conclusions are given in Section
\ref{Conclusion}.

\section{Method}
\label{Method}

We analyse the structural properties of dwarf galaxies
during minor mergers with dark satellites for a suite of hydrodynamical simulations
recently presented in \citet{SHS16}. The simulations span a range
of initial conditions for the dwarf galaxy, its satellite and a variety of 
orbital configurations for the interaction. They were performed using the OWLS
\citep{Schayeetal2010} version of Gadget-3 \citep[based on][]{SpYW01,
  Sp05} with implementations for star formation and feedback as
described in \citet{SDV08, DVS08}.

 \begin{table}[\textwidth, h!]
      \caption{\label{orbitparm} Parameters for the orbits}
         $$ 
         \begin{array}{llrcr}
            \hline
            \noalign{\smallskip}
            \mathrm{name} & \displaystyle\frac{v_r}{v_{\mathrm{vir_{\rm main}}}}  & \displaystyle\frac{v_t}{v_{\mathrm{vir_{\rm main}}}} & r_{\mathrm{apo}}/r_{\mathrm{peri}}\tablefootmark{a} & \mathrm{inclinations} \\    
            \noalign{\smallskip}
            \hline
            \hline
            \noalign{\smallskip}
            \noalign{\smallskip}
            \mathrm{standard} & 0;-0.08 & 0.06 &  \sim40  & 0;30\\
            \mathrm{wide-inclined} & 0;-0.08 & 0.64 & 4 & 10 \\
            \mathrm{wide} & 0;-0.08 & 0.86 & 2 & 0 \\
            \mathrm{intermediate} & 0;-0.08 & 0.5 & 6 & 0 \\
            \noalign{\smallskip}
            \hline
            \hline
         \end{array}
     $$ 
     \tablefoot{
       \tablefoottext{a}{The apo-to-peri ratio is an average for the first pericentric passage for different main and satellite masses and inclinations.
       }
     }   \end{table}

\begin{figure*}
\includegraphics[width=\textwidth]{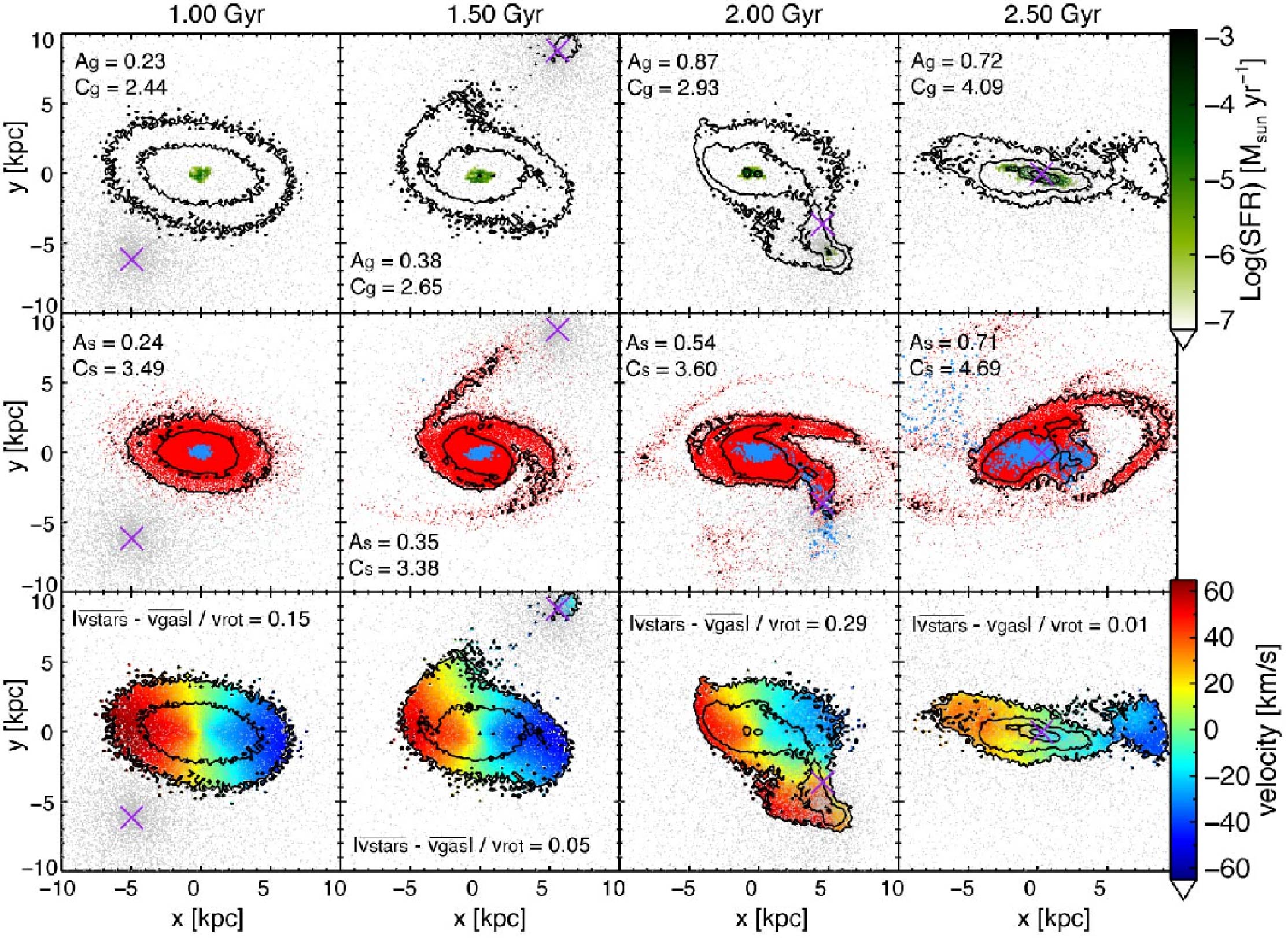}
\caption{\label{merger-lessradial-highC} Evolution of the model-A
  dwarf galaxy with $R_g= 4 R_d$, merging with a 20\% mass ratio dark
  satellite with $c=25$ on a co-planar relatively radial orbit (the intermediate orbit in Table \ref{orbitparm}). All
  figures show an inclined view of the disk, 60 deg from face-on. The top
  row shows the gas in the disk above a certain column density (contours at $0.4$, $1$, $4$, and
  $10\times 10^{20}\ \mathrm{N}\ \mathrm{cm}^{-2}$) with the
  star-forming gas highlighted in green (see colourbar for SFR
  values). The central row panels show the old stars in red, and those
  newly formed in blue, along with two surface brightness contours of
  $25$ mag/arcsec$^2$ and $28$ mag/arcsec$^2$, obtained assuming an
  $M/L = 0.5$ for all stellar particles. The bottom panels show the
  gas contours with the gas velocity maps. In all panels the satellite
  is shown in grey ($5\%$ of the particles are plotted), with the
  purple cross denoting its centre of mass. The insets indicate the
  values of asymmetry, concentration and difference in average
  velocity between stars and gas computed as described in
  Sec.~\ref{Quantitative}.}
\end{figure*}

The host dwarf galaxy consists of a dark matter halo, a stellar disk and a
(generally more radially extended) gaseous disk. Both the stellar and gaseous
disks follow an exponential surface density profile with radius, while
the vertical distribution of the gas is determined by requiring
hydrostatic equilibrium, and assuming an effective equation of state
of the multiphase interstellar medium (ISM) model by \citet{SDV08, DVS08}.  Star
formation occurs following the Kennicutt-Schmidt relation when the density of the gas is above a threshold of $0.1\, \mathrm{cm}^{-3}$, while at lower densities the gas follows an isothermal
equation of state (see \citet{SDV08} for more details). Feedback and stellar winds are included such that
the systems, when evolved in isolation, are self-regulating over the
timescale of the simulations. Our ISM model results in a more regular and spatially extended star-formation activity
than when using higher density thresholds for star formation and models that resolve the multi-phase
structure of the gas \citep[e.g.][]{Governatoetal2010, Hopkinsetal2011, Huetal2016}. 
However, although we expect the results to change quantitatively when using a different ISM model, 
the main trends identified in our merger simulations should remain. 
Indeed, the more modulated star formation history naturally associated to our model
suggests that the results presented here represent safe lower-limits to the star formation
events expected in more bursty prescriptions.
\begin{figure*} 
\includegraphics[width=\textwidth]{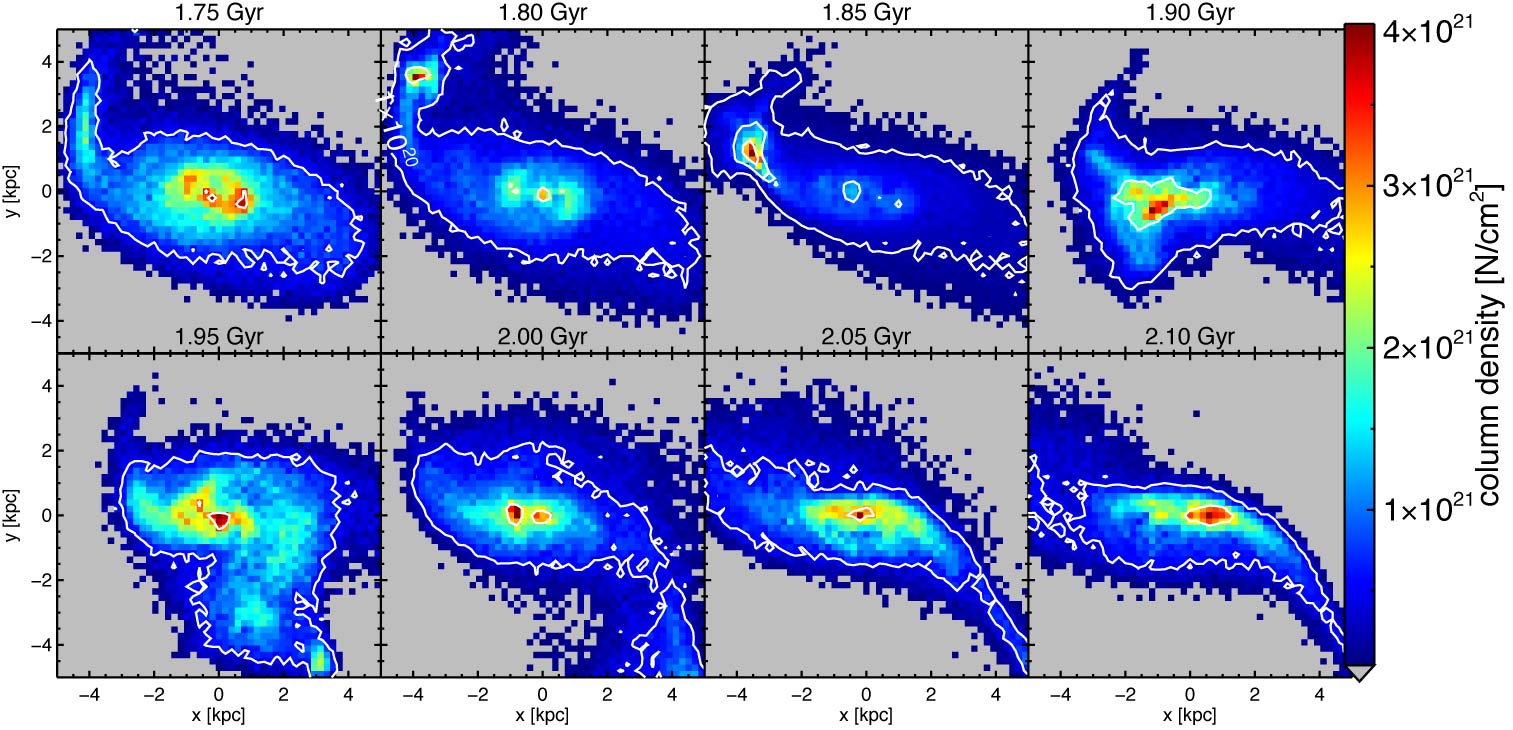}
\caption{\label{merger-lessradial-highC-gasinner} Evolution of the
  inner parts of the dwarf galaxy shown in
  Fig. \ref{merger-lessradial-highC} around the time the satellite
  merges with the host. The threshold is $0.4 \times 10^{20}\ \mathrm{N}\ \mathrm{cm}^{-2}$ and the contours indicate $1$, $4$, and $10\times
  10^{20}\ \mathrm{N}\ \mathrm{cm}^{-2}$; see colourbar for the 
  relative values.}
\end{figure*}

The satellite is a dark subhalo (no baryons) that follows an
NFW-profile with two different concentrations ($c \sim 15$--$18$ (based on the $M_{\mathrm{vir}}-c$ relation \citep{MunozCuartasetal2011}), $c=25$). In most of
the simulations, it has an initial mass of $20\%$ of that of the dwarf
galaxy's halo, but we also consider $5\%$ and $10\%$ mass ratios. Table \ref{hostparm} gives the virial radius of the dwarf galaxies and the mass of the satellites with respect to the virial mass of the dwarf galaxies. The satellite is
typically placed on a fairly radial orbit  (Table \ref{orbitparm} gives the general orbit parameters) with different inclinations
and is launched close to its apocentre (with $v_r/v_{\mathrm{vir}_{\mathrm{main}}} \sim 0$) at $0.67 r_{\mathrm{vir}}$ of the  host.

For the numerical parameters, we use $1 \times 10^6$ particles for the dwarf's dark matter halo, a 
softening length $\epsilon_{\mathrm{halo}}=0.025\ \mathrm{kpc}$, $2\times10^5$ particles in baryonic mass, 
divided among the stellar and gas disks according to the gas fraction $f_g = M_{\mathrm{gas}}/(M_{\mathrm{gas}}+M_{\star})$, 
with softening length $\epsilon_{\mathrm{bar}}=0.008\ \mathrm{kpc}$. The satellite is modelled with 
$1 \times 10^5$ particles that have a softening length
$\epsilon_{\mathrm{sat}} = 0.016\ \mathrm{kpc}$.

Table \ref{hostparm} contains the parameters for the dwarf galaxy and the ratio of the initial virial mass of the satellite to the dwarf galaxy virial mass.
The particle mass for the dark satellite is in between the particle masses of the main dark matter halo and the baryons to avoid numerical effects in interactions with both the halo and the disk.

We will focus mostly on one of the simulated dwarfs, which we refer to
as model-A \citep{SHS16} for which we explore ranges in the extend of the gas disk with respect to the stellar disk, 
the satellite mass, and the satellite orbit. We will also report results for smaller
mass systems with a range of disk thickness and halo concentrations, models B, C, and D in Table \ref{hostparm} (see \citet{SHS16} for more detailed information).

\section{Tidal effects}
\label{Tidal}

As an example, we present in Fig. \ref{merger-lessradial-highC} the
evolution of the model-A dwarf as it experiences a 20\% minor merger. In
this example the dwarf has initially a very extended gas disk, with
scale length $R_g = 4 R_d$. This set-up is motivated by observations
showing that gas may spread out much farther than the stars \citep[see e.g.][and references theirin]{Begumetal2008FIGGS, Filhoetal2015}.

The satellite in Figure \ref{merger-lessradial-highC} has a high
concentration ($c=25$) and is launched from apocentre at a
  distance of $\sim 51$ kpc on a fairly radial orbit with tangential velocity $v_t
= 0.5 v_{\mathrm{vir}}$. During close passages to the disk the satellite (marked with a cross)
induces large tidal tails in both the gas and stars, as shown in the
second column of this figure.

During the second pericentric passage (third column of
Fig. \ref{merger-lessradial-highC}), the satellite meets up with a
gaseous tidal tail and causes a local overdensity where star formation
takes place (as can be seen from e.g. the newly born stars plotted in
blue in the middle panel). The star formation rate density in this
tidal structure is higher than in the centre so that the
brightest star-forming core at this point in time is actually located
more than 7 kpc from the centre. Intriguingly, such features are also
found in extremely metal-poor galaxies (XMP; galaxies with oxygen abundances smaller than a tenth of the solar value) \citep{Filhoetal2013}.

Fig. \ref{merger-lessradial-highC-gasinner} zooms into the gas column
densities in the inner parts of the dwarf galaxy around this
time. Note again the high gas densities in the tidal tails and how
they correlate with the position of the satellite. Also, and as expected,
the young star particles trace the motion of the satellite through the
disturbed dwarf galaxy (see the blue points in the middle row of
Fig. \ref{merger-lessradial-highC}). Since these dominate the light,
the associated local star-forming regions may well be the analogues of
what is seen in XMP galaxies \citep{Filhoetal2013,Filhoetal2015}. 

\begin{figure}
\includegraphics[width=0.5\textwidth]{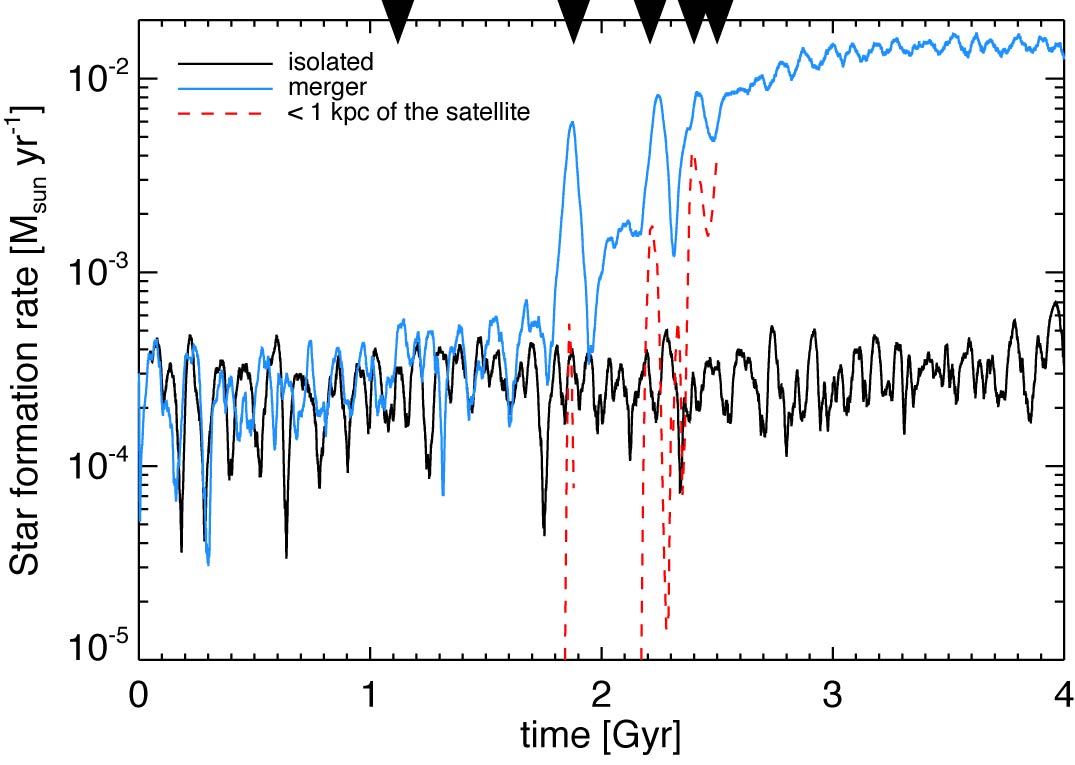}
\caption{\label{merger-lessradial-highC-SFR} The blue curve shows the evolution of the star
  formation rate of the dwarf galaxy shown in
  Figures \ref{merger-lessradial-highC} and
  \ref{merger-lessradial-highC-gasinner}, while the black curve is for the same
  model-A dwarf run in isolation. The star formation rate in the
  tidal tail within 1 kpc of the position of the satellite is shown
  with the red dashed curve, and constitutes a significant fraction of the total SFR in
  the system around the time of the merger. The satellite is completely disrupted after 2.5 Gyr, but
  its effects on the SFR are longlasting. The pericentre passages of the satellite are indicated by the
  black arrows.}
\end{figure}

Figure \ref{merger-lessradial-highC-SFR} shows the star formation
rate (SFR) during the encounter. The blue curve shows that the total SFR has
pronounced peaks during the pericentric passages of the satellite.
Interestingly, during the later passages, a significant fraction ($4 \times 10^{-3}\ M_{\sun}\ \mathrm{yr}^{-1}$) of the total
star formation takes place in the tidal tail at the location of the
satellite (red curve). Subsequently the total SFR increases, due
to the gas that is channeled to the centre, and reaches a plateau
around a value that is more than a factor 10 higher than for the equivalent dwarf 
in isolation. 

In summary, besides the characteristic starburst, signatures of the
merger can be found in the morphology of the old stellar
disk, the distribution and morphology of the young stellar population
that is formed during the encounter, and in the morphology and
kinematics of the gas. The gaseous and stellar disks show
distinguishable effects both in the outskirts and in the inner
parts. Interestingly, in the simulated system shown here star-forming regions outside
the centre pinpoint the location of the merging dark satellite.

\section{Quantitative measures of structural properties}
\label{Quantitative}

Although clear effects can be seen in the simulated dwarf galaxy which are due
to the minor merger, it is important to quantify these in order to
make comparisons to observations. A variety of quantitative structural descriptions
of the morphology and kinematics of galaxies have been put forward in
the literature. Morphologically the structure is often characterised
by the \emph{CAS} \citep[concentration, asymmetry, and smoothness;
][]{Conselice2003} and \emph{GM} \citep[Gini and M20;
][]{Lotzetal2004} indicators. Additionally for dwarf irregulars and
BCDs, the difference in central surface brightness obtained by fitting the inner and
outer regions is also used \citep{Hunteretal2006,
  Papaderosetal2008}. For XMP galaxies, the often used indicators
include differences in the average velocity, and in the position
angles of the HI gas and stellar components \citep{Filhoetal2015,
  Filhoetal2013}. We apply the morphological indicators to our
simulations and present the results in Sect. \ref{morphparm}, while we
focus on the results of the kinematic indicators of the stellar and
gas disks in Sect. \ref{kinparm}.

\subsection{Morphological parameters}
\label{morphparm}

We estimate the morphological \emph{CAS} and \emph{GM} parameters,
including modifications by \citet{Lellietal2014c, Holwerdaetal2011II},
on a grid with initial size of 20 by 20 kpc which is cropped to the
regions above a fixed threshold, and a default bin size of $0.2$
kpc. The thresholds adopted are close to those reported for observational studies in the literature, $N_{HI} > 4 \times 10^{19}\ \rm{ cm}^2$ for the surface density of neutral gas and $\mu_V < 28\ \rm{mag}/\rm{arcsec}^2$ for the V-band magnitude, respectively. 
This means that in practice each bin holds at least $12$ gas particles. 
For the stars, assuming an average $M/L=0.5$, appropriate for the V-band, each bin holds at least $4$ stellar particles.
 The calculations are done on the stellar densities though (in $M_{\odot}/\rm{kpc}^2$).

For many of the indicators it is necessary to define
the centre of the system. This is done by fitting a 2-dimensional
Gaussian to the projected density (although our results are robust to the centering method used), with the threshold values
described above. We have tested the effect of different thresholds
($N_{HI} > 10^{19}\ \rm{ cm}^2$, $N_{HI} > 10^{20}\ \rm{ cm}^2$, $\mu_V < 26\ 
\rm{mag}/\rm{arcsec}^2$, and $\mu_V < 30\ 
\rm{mag}/\rm{arcsec}^2$), and assumptions regarding the mass-to-light
ratios for the newly formed and original stellar populations, and found
that when a sufficient area of the system is visible (as smaller
systems can mostly disappear below the thresholds) the numerical
values for the morphological parameters can change but the trends stay intact.

\subsubsection{Definitions}

We now describe in detail the different morphological indicators we use in our analysis. 
\begin{itemize}
\item \emph{Concentration} \\
This describes the distribution of light over the image:
\begin{equation}
C=5 \rm{ log } \left( r_{80}/r_{20} \right) 
\end{equation}
where $r_{80}$ and $r_{20}$ are the radii which contain 80\% and 20\%
of the total light \citep{Conselice2003}, where for a purely
exponential profile $C=2.7$, and for a de Vaucouleurs profile
$C=5.2$. We note that since the projected surface brightness and gas
column densities are computed on a grid, we determine a slightly
coarse value of $C$.
\\

\item \emph{Asymmetry} \\
This describes the relative difference in intensity when the image is rotated 180 degrees:
\begin{equation}
A = \frac{\sum_{i,j} |I(i,j) - I_{180}(i,j)|}{\sum_{i,j} I(i,j)} 
\end{equation}
where $I(i,j)$ is the intensity of the pixel $(i,j)$ \citep{Conselice2003}.
With this definition, $0 < A < 2$. \\

\item \emph{Outer Asymmetry} \\
To give more weight to the outer parts, \citet{Lellietal2014c} have defined an outer asymmetry parameter as: 
\begin{equation}
OA = \frac{1}{N} \sum_{i,j} \frac{|I(i,j) - I_{180}(i,j)|}{|I(i,j)+I_{180}(i,j)|} 
\end{equation}
\citep{Lellietal2014c}, where we define $N$ as the number of the pixels with $|I(i,j)+I_{180}(i,j)| > 0$. However, 
this outer asymmetry (OA) indicator is more susceptible to noise. 

Both for the $A$ and $OA$ parameters the detectability of asymmetries in the
outskirts depend greatly on the surface brightness or column density
thresholds, especially for low mass and low surface brightness
galaxies.
\\

\item \emph{$M_{20}$} \\
  This parameter is a relative second order moment of the 20\%
  brightest pixels and was originally introduced as an alternative to
  the concentration parameter:
\begin{equation}
M_{20}= \mathrm{log} \left( \frac{\it{\sum^k_i M_i}}{\it{M}_{\rm{tot}}} \right) 
\end{equation}
where $\sum^k_i I_i < 0.2 I_{\rm{tot}}$ and $M_i =
I_i[(x_i-x_c)^2+(y_i-y_c)^2]$ \citep{Lotzetal2004}.  Its advantage
compared to the concentration is that there is no assumption of
circular symmetry and that it is more sensitive to phenomena like
multiple nuclei that are thought to be common in merging, or
post-merging, systems.
\\

\item \emph{Gini coefficient} \\
This statistic originally used in economics to describe the distribution of wealth within a society, was adapted to galaxy morphology by \citet{Abrahametal2003}. It correlates with concentration but does not assume circular symmetry. 
We use the Gini coefficient based on the second intensity moment weighted by position:
\begin{equation}
G(M)=\frac{1}{2\bar{M} N(N-1)}\sum_{i,j} |M_i-M_j| 
\end{equation}
where again $M_i = I_i[(x_i-x_c)^2+(y_i-y_c)^2]$, and $\bar{M}$ denotes the mean of $M_i$ over all $N$ pixels \citep{Lotzetal2004,Holwerdaetal2011II}.
$G(M)$ is larger when the brightest pixels are farther from the centre.
We only consider the pixels above the threshold in this calculation, which tends to lower the values of $G(M)$ compared to including the background pixels. \\

\item \emph{Excess central surface brightness: $|\mu_{0,HSB} - \mu_{0, LSB}|$} \\
We compute this by taking the difference in the central bin's surface brightness 
obtained from exponential fits to the inner and outer parts of the system.
\end{itemize}

\subsubsection{Results}
\begin{figure}
\includegraphics[width=0.5\textwidth]{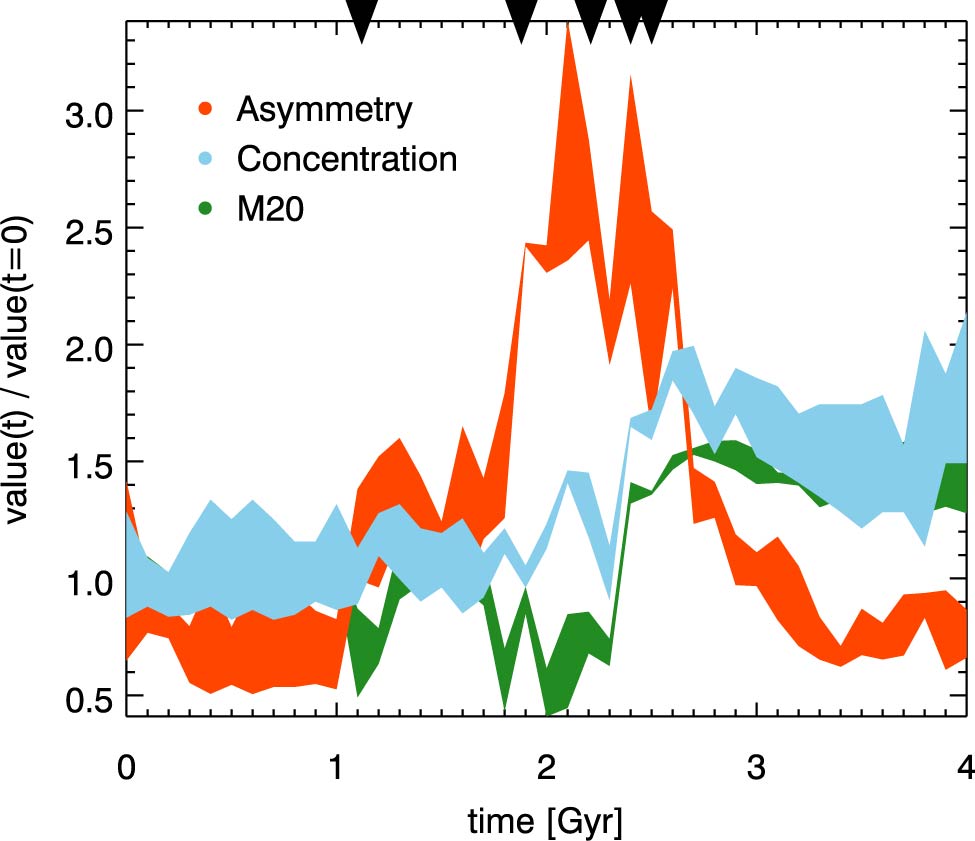}
\caption{\label{A-parm} The distribution of the asymmetry $A$ (red),
  concentration $C$ (blue), and $M_{20}$ (green) values for the gaseous disk for 5 random
  inclinations in time intervals of 0.1 Gyr for the model-A dwarf
  during the minor merger shown in
  Fig. \ref{merger-lessradial-highC}. These quantities have been
  normalised to their median (over all inclinations) initial
  value. The pericentre passages of the satellite are indicated by the
  black arrows.}
\end{figure}

Fig. \ref{A-parm} shows the evolution of three morphological
indicators: asymmetry, concentration and $M_{20}$, for the
gas distributions in the system shown in
Fig.~\ref{merger-lessradial-highC}, for five different random
inclinations. This figure evidences that the indicators have a strong
time dependence as the merger occurs, and that each evolves quite differently with
time. For example, the asymmetry (in red) increases with time reaching
a peak value when the satellite fully merges (with less than 100 particles, $0.1 \%$ of the initial satellite mass, of the satellite still gravitationally self-bound), as a consequence of the
extended tidal tails clearly seen in
Fig.~\ref{merger-lessradial-highC}, and decreases strongly
afterwards. The concentration (in blue) also increases significantly
around the time of the merger but it remains high afterwards, as a
result of the strong central influx of gas. On the other hand, the
$M_{20}$ (in green) depicts an oscillatory behaviour with peaks that
roughtly coincide with each pericentre passage of the
satellite. Because $M_{20}$ is negative, and in this figure it has
been normalised to the initial value, these peaks actually imply that
the 20\% brightest pixels are more centrally
concentrated, with the dips indicating high gas densities at larger radii, suggesting that star formation occurs at
larger distances. The plateau value at late times reflects the strong centrally concentrated
sustained enhancement in gas density. The parameters describing the stellar distribution follow similar trends.

\begin{figure*}
\includegraphics[width=\textwidth]{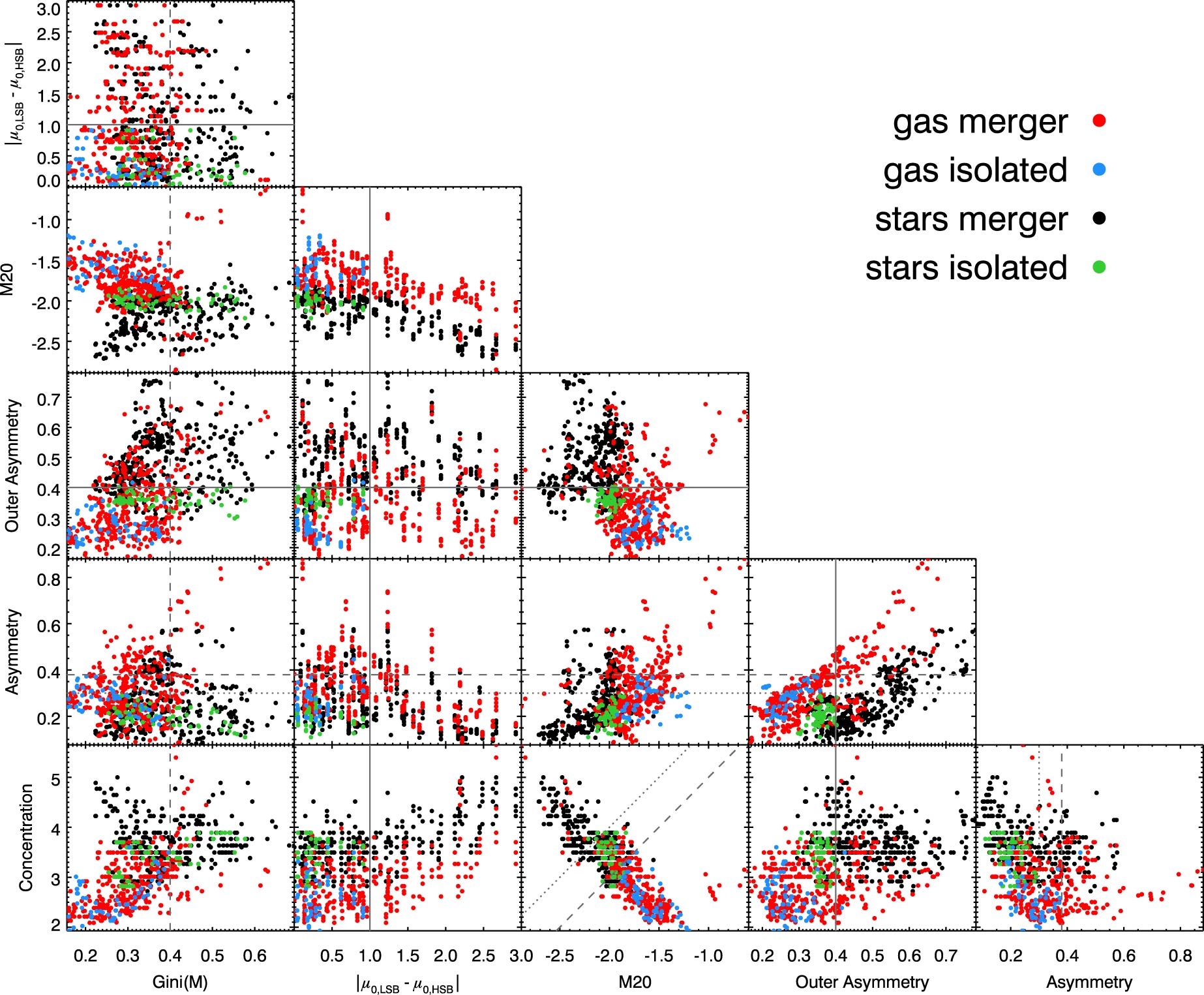}
\caption{\label{CAS-H1H2H4} The concentration $C$, asymmetry $A$,
  outer asymmetry $OA$, $M_{20}$, Gini($M$), and excess central surface
  brightness $|\mu_{0,HSB} - \mu_{0, LSB}|$, for the gas in merger
  simulations (red) and in isolation (blue), and for the stars in
  merger simulations (black) and in isolation (green). The simulations
  shown correspond to the model-A dwarf, and encompass 14 different
  runs with varying satellite masses, halo concentrations, orbits,
  and radial extend of the gas disks with respect of the stellar
  disks (with $R_g = R_d$, $2 R_d$, or $4 R_d$) as
  described in Sect. \ref{Method}. For more details on the
  simulations, see \citet{SHS16}. The dotted, dashed and solid lines
  indicate regions where the isolated and merger systems are well separated
  for the stars, for the gas or for both, respectively.}
\end{figure*}

Fig. \ref{CAS-H1H2H4} shows the distribution of photometric indicators
for all the merger simulations we have carried out with the model-A
dwarf. These simulations include a range of different orbits
(orbital inclinations and eccentricities), concentrations and masses
for the satellite, and varying extents of the gaseous disk. For each
simulation the parameters are calculated initially (i.e. in isolation)
and at 1, 2, 3, and 4 Gyr and at five random inclination angles for
each point in time.

This figure shows that isolated systems (blue for gas, and green for
stars) tend to occupy small regions of parameter space, whereas for
mergers (red for gas and black for stars) a broad range of parameter
values appears to be plausible. At face value, there is no parameter
(combination) for which mergers and isolated systems can be fully
separated. This might not be unexpected given the time variability of the
parameters. Furthermore, cases in which the effects of the merger on
the gas and stellar disks are small (e.g. if the satellite sinks in
very slowly or has too low mass, or for specific viewing angles), 
will be hard to disentangle from systems in isolation.

Most isolated systems have low values for concentration, asymmetry,
outer asymmetry, and $G(M)$. On the other hand, for the mergers, the asymmetry
parameters for both the gas and the stars spread over a much larger
range. Also the outliers in $M_{20}$ correspond to
merging systems.  The difference in central surface brightness can
reach up to to 3 magnitudes/arcsec$^2$ for merging systems, but is
smaller than 1 magnitude/arcsec$^2$ for all isolated cases. 

It is therefore easier to demark regions populated by isolated systems
in the parameter subspaces plotted in Fig. \ref{CAS-H1H2H4}. For example $A < 0.38$ for
the gas, $A < 0.3$ for the stars, $OA < 0.4$ for gas and stars, $G(M)
< 0.4$ for the gas, and a relation $C \lessapprox 2 M_{20} + 7$ for
the gas and $C \lessapprox 2 M_{20} + 8$ for the stars. These regions
are indicated by grey lines in the figure. Interestingly
we find $G(M) > 0.4$ for the HI component of merging systems, while $G(M) > 0.6$  has been put
forward by \citet{Holwerdaetal2011II}, and $A > 0.4$
has been used for the stellar component in major mergers
 \citep{Conselice2003}. 

In Fig. \ref{CAS-H1H2H4} we have focused on the model-A dwarf, a
relatively massive system with $M_{\star} = 1.4 \times 10^{8}
M_{\sun}$, and demonstrated that it may be possible to disentangle
partly mergers from isolated systems. However, for lower mass dwarfs,
with initial $M_{\star} = 4.4 - 27 \times 10^6 M_{\sun}$, the
morphological parameters of either isolated and merging systems
strongly overlap. Although a 20\% merger can cause irregular features
in the gas and stellar distributions (ideally resulting in higher asymmetry and outer asymmetry values),
often these features are not strong
enough (given reasonable thresholds) to be clearly identified by the
\emph{CAS} or \emph{GM} indicators as being different from irregular
gas densities and patchy star formation that may happen in isolation
as well. Therefore, such morphological indicators are not useful to
identify merger candidates in the case of low mass dwarfs.

\subsubsection{Comparison to observations}

\citet{Lotzetal2004} have estimated the asymmetry, concentration, and
$M_{20}$ parameters for 22 systems from a sample of isolated dwarf
irregular galaxies observed in the B-band by
\citet{VanZee2000,VanZee2001}. Many of these systems are brighter than
those in our simulations and seem to be more clumpy. Although the
range of concentrations is similar ($2.39 < C <4.17$), the
values for $M_{20}$ are higher ($-1.79 < M_{20} < -0.70$)
than we find for the stellar components even in isolation. This
implies a smoother distribution in the simulations, and this could be
the result of an initial smooth set up as well as to the absence of H$_2$
or metal-line cooling in the simulations which could induce a patchier
star formation. On the other hand, the asymmetry values of these observed late-type dwarf galaxies are in the
range of $0.01 < A < 0.45$ \citep{Conselice2003}. We note that the observed sample is not selected as likely merger remnants but because they are faint, gas-rich, and isolated. The asymmetry values are consistent with what we find for the stellar components of isolated systems.
\begin{figure*}
\includegraphics[width=0.9\textwidth]{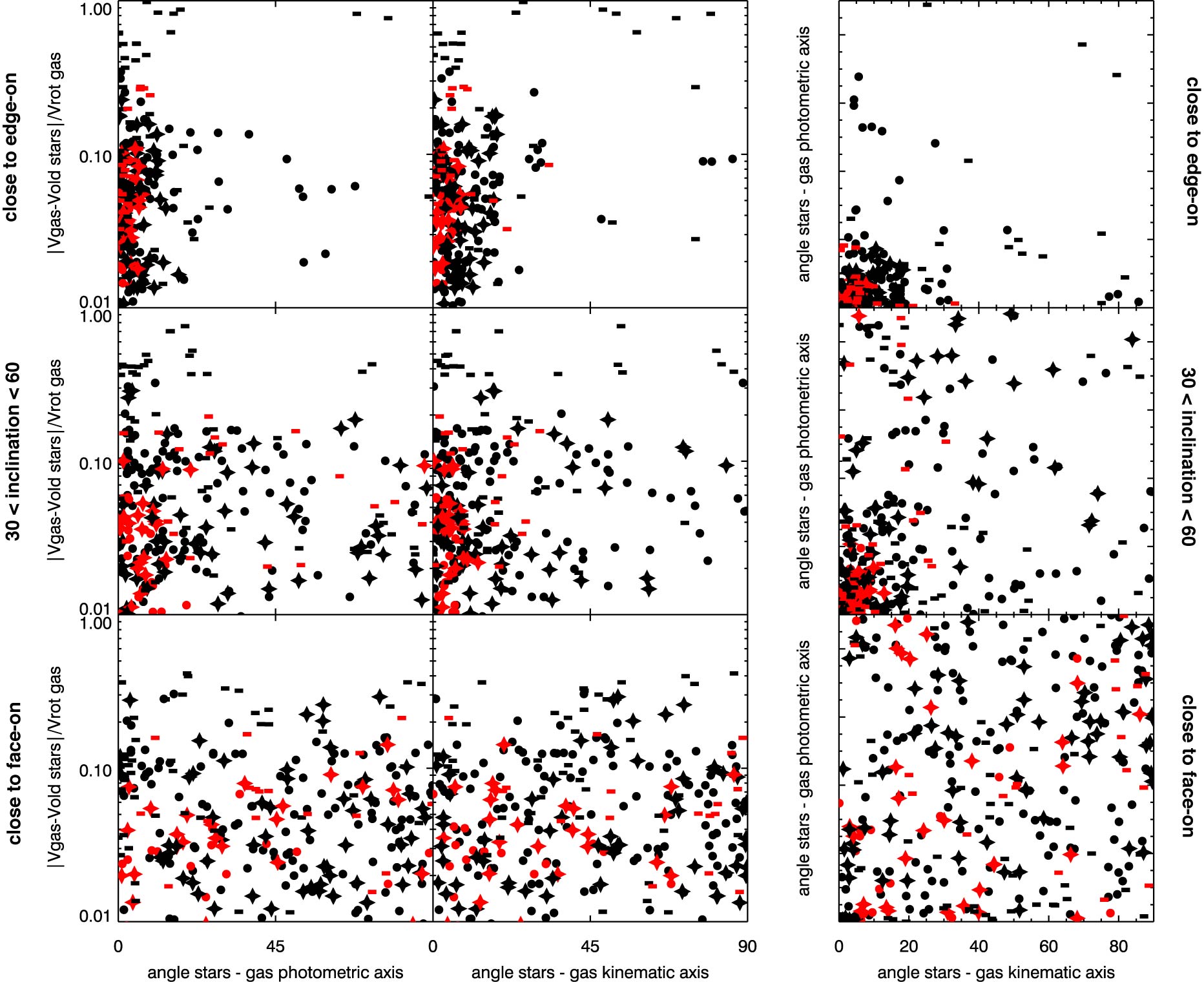}
\caption{\label{kin-H1H2H4} Average velocity differences between the
  gas and the stellar disk normalised to the average observed
  rotational velocity of the gas, difference in position angle between the projected
  distributions of the gaseous and stellar disks, and
  between the projected stellar disk and the kinematic axis of the
  gas, for merger simulations (black) and for systems in isolation
  (red). The different rows correspond to different ranges of viewing
  angles. In this figure we have included the
  model-A dwarf with $M_\star = 1.4 \times 10^{8}\ M_{\sun}$ (solid
  circles), as well as values for systems with $M_\star = 4.4 \times
  10^6\ M_{\sun}$ (dashes) and with $M_\star = 1.1 - 2.7 \times 10^7\
  M_{\sun}$ (diamonds).}
\end{figure*}

\citet{Lellietal2014c} have described the gas outer asymmetry for 18
starburst dwarf galaxies and for a control sample of 17 dwarf
irregular galaxies from the VLA-ANGST survey \citep{Ottetal2012}. The
outer asymmetry values of the observed starburst systems are in the range $OA =
0.42 - 0.77$, with a median value of $\sim 0.6$, that is similar to the values we find in our
merger simulations. On the other hand, all the simulated dwarf galaxies in isolation have 
outer asymmetries lower than $0.4$, and hence are more comparable to those in the dwarf irregular 
sample, which typically have $ OA \sim 0.3 - 0.5$.

From these comparisons, we may conclude that both the stellar and gas components of
dwarf irregular galaxies have similar parameter distributions to the
simulated dwarfs in isolation. Furthermore, the outer asymmetries seen
in the gas in observations of starburst dwarf galaxies agree with those of
interacting simulated dwarfs.

\subsection{Kinematic parameters}
\label{kinparm}

Besides morphology, kinematics can also encode information about past
merger events.  For example, in our merger simulations the
3-dimensional direction of the total angular momentum vectors of the
gas and of the stars can differ significantly, and up to 60
degrees, while for the isolated simulated dwarfs the difference is
$<5$ degrees. However, angular momenta cannot be directly measured from
observations and so we discuss below some of the kinematic indicators
that may be used instead.

\begin{itemize}
\item \emph{Difference in average velocity between gas and stars} \\
  This has been found to be quite large for a number of extremely
  metal-poor (XMP) galaxies \citep{Filhoetal2013}. In our simulations
  the average line-of-sight velocities are mass-weighted and computed for all
  particles within a bin with surface brightness or column density
  above the thresholds. We compare this difference to the ``maximum''
  rotational velocity defined as $\frac{1}{2}
  (|\mathrm{max}(v_{\mathrm{proj}})| +
  |\mathrm{min}(v_{\mathrm{proj}})|)$, where these stem from the 
  projected gas velocities within the observed region.
  \\

\item \emph{Misalignment between stars and gas} \\
  A relatively straightforward measurement consists in comparing the orientation of the major axis of
  the surface brightness to that of the projected gas distribution.  These are
  computed by fitting a 2D-gaussian to these projected distributions.

  Since measuring velocity fields for stars is challenging, in
  general it will not be possible to estimate the 
  misalignment between the kinematic axes of stars and of gas. Therefore instead, we compare the orientation of the photometric major
  axis of the stars to the gas kinematic major axis which can easily
  determined observationally from HI velocity maps.

 The gas kinematic major axis is determined in our simulations using the line connecting
  the maximum and minimum velocities observed. To obtain an estimate
  of the uncertainty in the orientation we compute the kinematic axis
  50 times, each time using two randomly chosen values amongst those ranked in the top 10\% as maximum and
  minimum. From this random sampling we
  estimate an uncertainty of $9.8$ deg, for the merging systems (lower for isolated systems). However, this estimate depends
  strongly on the amplitude of the velocity field, for example for systems
  close to face on, the uncertainty can be as large as $\sim 54$
  degrees.
\end{itemize}

Figure \ref{kin-H1H2H4} presents the results for all our simulated
systems.  We have separated the analysis according to the projected
inclination because this has a significant impact on the ability to
separate isolated from merging systems.

For low mass systems (indicated by the dashes), we find the largest
average velocity differences between stars and gas in mergers, while for all isolated systems, independently of their mass, $\Delta = |v_{\rm gas} - v_{\rm stars}| \lesssim
0.1 v_{\mathrm{rot}}$, and this appears to be relatively robust to
inclination effects. Inspection of the simulations shows that the
largest amplitude is reached around the time the satellite reaches the
disk, that is around the first pericentre passages. 

The velocity differences in our simulations are typically smaller (a few
km/s) than those observed for XMP dwarfs by \citet{Filhoetal2013}
(These authors disregard offsets smaller than 10 km/s because of
uncertainties and their expectation that HI velocity dispersions are
$\sim 10$ km/s for dwarf galaxies). However, their normalised velocity differences (the difference between the average HI line-of-sight velocity and the average velocity of the stars divided by the full width at half max of the HI line, $w_{50}$),
are $0 \lesssim \Delta_{\mathrm{HI}}/w_{50} \lesssim 1$, and
hence consistent with those in our simulations.

Comparison of the different rows in Figure \ref{kin-H1H2H4} directly
shows that the effects of inclination are important. Especially for
nearly face-on systems, the separation between mergers and isolated
dwarfs is not straightforward. This is entirely due to the large
uncertainties in the determination of the orientation of the photometric and kinematic axes. For
example, isolated systems have close to circular spatial
distributions, so that major and minor axes directions are hard to
define. Furthermore, the line-of-sight velocities are typically small in this case
and so also the rotation axis is not well constrained. This leads to more
scatter in these distributions.

For other inclinations, the isolated systems tend to be clustered
around small average velocity differences, and small misalignments.
In other words, mergers are clearly more likely to have misaligned
stellar photometric and gas photometric or kinematic major axes. The
lack of correlation seen in the bottom left panel of
Fig. \ref{kin-H1H2H4} is a result of the misalignment between the 
photometric and kinematic axes for the gas in the case of mergers.

\begin{figure}
\includegraphics[width=0.5\textwidth]{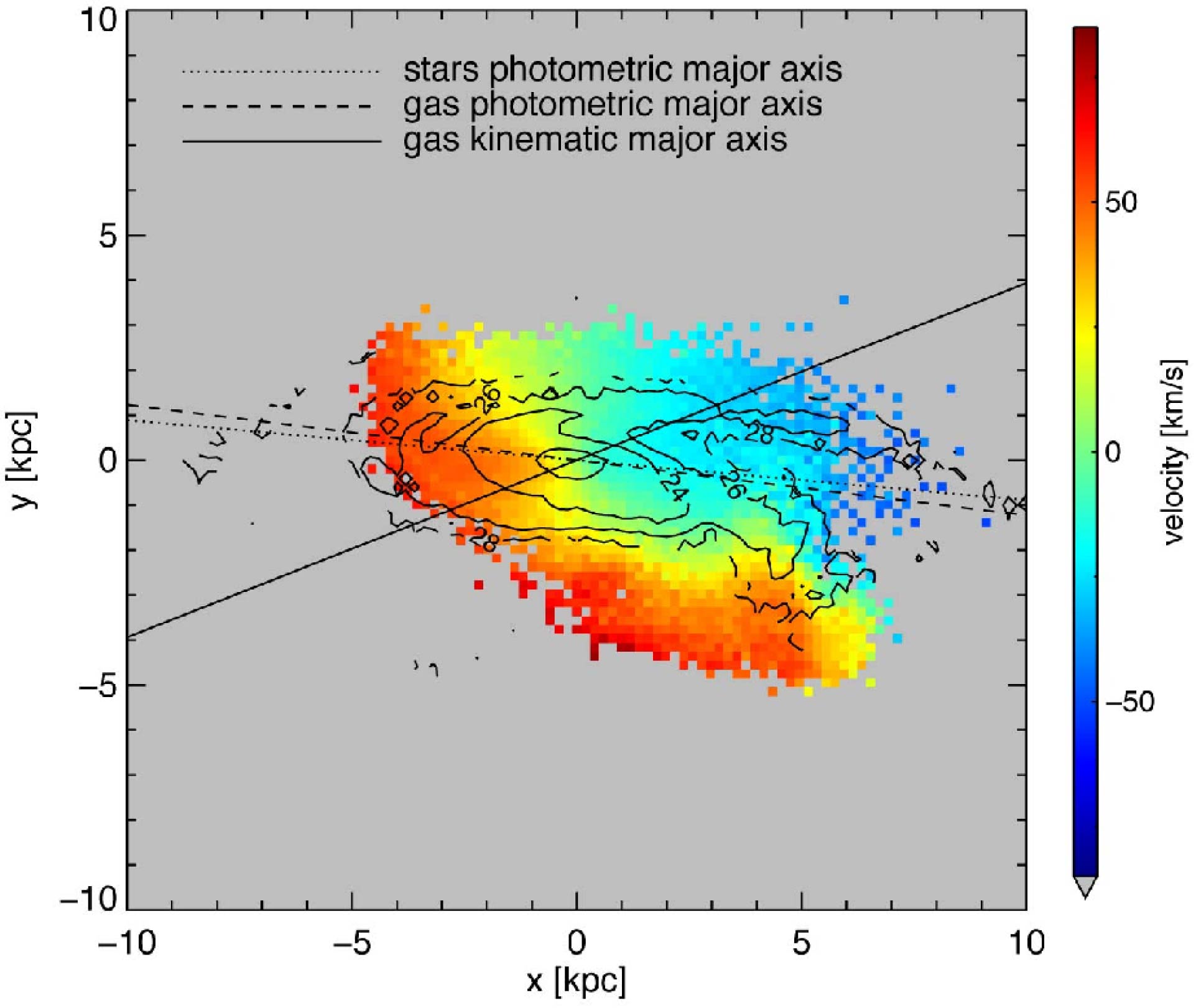}
\caption{\label{A-2Gyr-kinopt} Gas line-of-sight velocity field with the stellar surface brightness overplotted as contours for the model-A dwarf system during the minor merger shown in Fig. \ref{merger-lessradial-highC} at $t=2$ Gyr, but now for an inclination angle of $i=72.6$ deg from face-on. The kinematic axis for the gas (solid line) and the major axis of the surface brightness distribution (dotted line) are misaligned. The fitted profiles to the stellar and gas (dashed line) distributions have similar orientations, and this is  because they are dominated by the behaviour in the central regions.} 
\end{figure}

Fig. \ref{A-2Gyr-kinopt} provides a visual impression of a projection
where the gas kinematic and the stellar distribution major
axes are misaligned for the system from
Fig. \ref{merger-lessradial-highC} at 2 Gyr seen for an inclination of
$72.6$ degrees. The gas column density distribution and stellar
surface brightness distribution have roughly the same orientation (the
misalignment angle is $\sim 2$ degrees), but for both the orientation
in the inner regions is different from that in the outer parts. The
gas kinematic axis however has a significantly different major axis
orientation, offset by $\sim 28$ degrees.

\section{Some intriguing cases}
\label{Discussion}

So far we have focused on general trends followed by our morphological and kinematic indicators, and especially on the
differences between isolated and merger systems to facilitate the
observational identification of dwarf galaxies undergoing a merger
when the secondary is not visible, in our case being a dark
satellite. We now make a rough comparison to a few intriguing cases
from the literature.

The system depicted in Fig. \ref{merger-lessradial-highC} shows a
distribution of gas and stars that shares characteristics with the
irregular dwarf galaxy IC10: a disturbed gas and stellar distribution
with multiple star-forming cores and an extended HI distribution with
plums and spurs with velocities that differ from that of the main gas disk
\citep[see for example][]{Ashleyetal2014}. On the other hand, the HII regions have a low metallicity \citep{Garnett1990} which has been
suggested as being due to the influx of fresh pristine gas from the
environment \citep{SanchezAlmeidaetal2014}. However, another interpretation is possible
since as we have seen the merger leads to an extended starburst
that is fueled from gas that was originally present in the outskirts
of the main system, and which presumably also had a lower
metallicity \citep[see also][]{SHS16}.

Another intriguing
system, though for different reasons, is the dwarf irregular galaxy NGC6822. In
addition to a disturbed gas and stellar distribution and a high rate
of recent star formation, this system has a star formation core
located very far from the centre. This outer star-forming region was
proposed to indicate the location of a companion system, also due to a
significant velocity offset \citep{deBlokWalter2000}, but this has
been discarded because no older stellar population has been found at
that location \citep{Cannonetal2012}. An interaction with a dark
substructure will however display exactly this signature: a star
formation region at a large distance without an underlying older population and a
metallicity similar to the main system. 

\section{Conclusions}
\label{Conclusion}

We have investigated the distribution of quantitative morphological
and kinematic parameters (often used to characterise interacting,
starburst, or peculiar systems), measured during a minor merger
between a dwarf galaxy and a dark satellite. For our system with $M_{\star} > 10^8\ M_{\sun}$ the very disturbed
morphologies for the gas and stellar distributions induced by the
merger are reflected most notably in asymmetry parameters during the
merger itself. A post-merger system however stands out the most in its
high values for concentration related parameters, such as $C$, $M_{20}$
and the Gini coefficient. This is the result of an increase in central
stellar and gas density due to gas being driven toward the centre by
tidal torques and causing a nuclear starburst episode, which can last several Gyrs.

Kinematic based parameters can be used to identify merger systems, for
example via the large differences between average projected gas and
stellar velocities. This works particularly well for smaller mass
systems ($M_{\star} < 2 \times 10^7\ M_{\sun}$), for which the morphological indicators fail. Misalignments
between the gas kinematic major axis and the stellar surface
brightness major axis are also useful, but can only be applied for
systems that are far from face-on. 

Although we still have to determine the smoking-gun that will allow
to determine that an interaction between a dwarf galaxy and a dark
satellite has taken place, in our simulations star-forming cores
located far from the centre actually seem to pinpoint the location of
the satellite. This could be the way to shed light on a missing satellite.

\begin{acknowledgements}
We are grateful to Claudio Dalla Vecchia, Joop Schaye,
Carlos Vera-Ciro, Alvaro Villalobos and Volker Springel for providing code.
AH acknowledges financial support from the European Research Council under
ERC-StG grant GALACTICA-240271 and the Netherlands Research Organisation
NWO for a Vici grant.
\end{acknowledgements}

\bibliographystyle{aa} 
\bibliography{TKS} 

\end{document}